\documentclass{nature}

\usepackage{color}
\usepackage{url}
\usepackage{graphicx}
\usepackage{aas_macros}
\usepackage{amssymb,amsmath}
\usepackage{multirow}

\renewcommand{\thefootnote}{\alph{footnote}}

\long\def\symbolfootnote[#1]#2{\begingroup%
\def\thefootnote{\fnsymbol{footnote}}\footnote[#1]{#2}\endgroup}

\title{A possible close supermassive black-hole binary in a quasar with optical periodicity}

\author{Matthew J. Graham$^{1}$, S. G. Djorgovski$^{1}$, Daniel Stern$^{2}$, Eilat Glikman$^{3}$, Andrew J. Drake$^{1}$, Ashish A. Mahabal$^{1}$, Ciro Donalek$^{1}$, Steve Larson$^{4}$  \& Eric Christensen$^{4}$}

\begin{document}

\maketitle

\noindent Published online by {\sl Nature} on 7 January 2015. DOI : 10.1038/nature14143

\begin{affiliations}
\item California Institute of Technology, 1200 E California Blvd, Pasadena, CA 91125, USA
\item Jet Propulsion Laboratory, California Institute of Technology, 4800 Oak Grove Drive, Pasadena, CA 91109, USA
\item Department of Physics, Middlebury College, Middlebury, VT 05753, USA	
\item University of Arizona, Department of Planetary Sciences, Lunar and Planetary Laboratory, 1629 E. University Blvd, Tucson, AZ 85721, USA
\end{affiliations}

\begin{abstract}
{
Quasars have long been known to be variable sources at all wavelengths. Their optical variability is stochastic, can be due to a variety of physical mechanisms, and is well-described statistically in terms of a damped random walk model\cite{2009ApJ...698..895K}. The recent availability of large collections of astronomical time series of flux measurements (light curves)\cite{2002AcA....52..397P,1993AcA....43..289U,2009PASP..121.1334R,2011AJ....142..190S} offers new data sets for a systematic exploration of quasar variability. Here we report on the detection of a strong, smooth periodic signal in the optical variability of the quasar PG 1302-102 with a mean observed period of 1,884 $\pm$ 88 days. It was identified in a search for periodic variability  in a data set of light curves for 247,000 known, spectroscopically confirmed quasars with a temporal baseline of $\sim9$ years. While the interpretation of this phenomenon is still uncertain, the most plausible mechanisms involve a binary system of
two supermassive black holes with a subparsec separation.  Such systems are an expected consequence of galaxy mergers and can provide important constraints on models of galaxy formation and evolution.} 
\end{abstract} 

Subparsec supermassive black-hole (SMBH) binary systems are not resolvable except possibly with long baseline radio interferometry. An alternative approach to their detection is through a modulated variability -- caused by, for example, perturbations in their accretion disks or precession of relativistic jets, if they are present (see Fig.~1). The best known candidate, OJ 287\cite{2008Natur.452..851V}, has shown a pair of outburst peaks every 12.2 years for at least the past century: this object can be interpreted as a secondary SMBH perturbing the accretion disk of the primary SMBH at regular intervals\cite{2011ApJ...729...33V}. Systematic searches for equivalent systems to date\cite{2013ApJ...777...44J,2013ApJ...775...49S,2011ApJ...738...20T} have attempted to identify them from broad-line velocity offsets in their optical and near-infrared spectra but cannot detect the closest pairs (with $\lesssim$0.1 pc separation). 

We applied a novel joint wavelet and autocorrelation function (ACF) based technique that identifies objects exhibiting strongly periodic behavior in their light curves (M.J.G. et al, manuscript in preparation) to the largest 
set of quasar time series currently available. These are drawn from the Catalina Real-time Transient Survey (CRTS)\cite{2009ApJ...696..870D,2010fym..confE..32D,2011BASI...39..387M}. PG 1302-102 (see Fig.~2) is the strongest periodic candidate out of twenty objects meeting the selection criteria: strong constant wavelet peak, strong ACF detection of periodic behaviour, sufficient temporal coverage for 1.5 or greater cycles at the detected period, and a phased light curve well-described by a sinusoid. For statistical comparison, we have also generated a simulated light curve for each known quasar based on a damped random walk (DRW) model, a standard statistical description of the optical variability of quasars\cite{2009ApJ...698..895K}, using the CRTS time sampling. We find that only one object from the simulated data of 247,000 quasars satisfies the same selection criteria, showing that the number of quasars selected is statistically significant and that strongly periodic behavior is not expected as an artifact of a DRW process.

PG 1302-102 has a median $V$-band magnitude of 15.0 and a redshift of 0.2784 which gives an absolute $V$-band magnitude of $M_V = -25.81$, assuming the 9-year WMAP cosmology\cite{2013ApJS..208...19H}. It is outside the footprint of the Sloan Digital Sky Survey but is associated with bright infrared and X-ray sources. It is also a very bright (720 mJy at 4.86 GHz), core-dominated flat spectrum radio source. Its optical/NIR spectrum (see Fig.~3) shows broad emission lines (H$\beta$, H$\alpha$, Pa-$\beta$, Pa-$\alpha$) with an inferred mass of $\log(M/M_{\odot}) = 8.3 - 9.4$ and the object appears to be radiating at or close to its theoretical Eddington limit ($\log(L/L_{Edd}) = 0$). The light curve for the quasar is well-fit by a sinusoid with an observed period of $1,884 \pm 88$ days (corresponding to a restframe period of $1,474 \pm 69$ days) and an amplitude of $\sim$0.14 mag. CRTS data (covering $\sim$1.8 cycles; that is, May 2005 to present day) is augmented by archival monitoring data\cite{1999MNRAS.309..803G,2000AJ....119..460E} available back to May 1993 giving a total of 4.1 cycles' coverage. This data is consistent with the behaviour seen in the past nine years of CRTS data, particularly with stochastic photometric variation imposed on a periodic signal. Further simulations show that the detection is statistically significant, with an observed signal 40 times the scatter from the mean.

As PG 1302-102 is bright and nearby, it has featured in a number of studies of quasars and their host galaxies. The radio and optical structure of the source is noted to be unusual. Hubble Space Telescope (HST) imaging\cite{1995ApJ...450..486B} shows that the quasar resides in a luminous elliptical host, as typical for radio-loud quasars\cite{1999MNRAS.308..377M}. There are also two companion galaxies that lie at projected distances of 3 and 6 kpc. Several features in radio images of PG 1302-102, such as the small radio core and sharp bends in the radio structure very close to the central source, correspond with features seen in the optical\cite{1994PASP..106..642H}. An interpretation is that the host galaxy is a fairly old merger, but that there might be more recent activity with the radio source just turning on and possible radio jets just emerging from the host galaxy. There may also be some indication of relativistic beaming connected with a jet. It should be noted that OJ 287 also exhibits a similar radio and optical morphology\cite{1996ApJ...464L..47B}.  

PG 1302-102 was spectroscopically monitored over a six-month period in 1990\cite{1992A&A...262...17J} and showed no detectable (greater than 5 -- 10\%) change in any component of its spectrum over that time. This lack of variation is not inconsistent with the $\sim$60 month period that we have identified. A SMBH binary may also exhibit double-peak broad line profiles in its spectrum for a small window of separation between the pair\cite{2010ApJ...725..249S} (although disk emission from an accretion disk around a single source may also 
produce the same effect\cite{2010Natur.463E...1G}). At closer distances, the two black holes dynamically affect the broad-line region clouds as a single complex entity producing single-peaked spectral lines with asymmetric line profiles. The Balmer and Paschen series spectral lines in PG 1302-102 do not show a double peak profile
but are consistently asymmetric (see Fig.~4). In particular, a small bump on the red wing of H$\beta$ has been reported\cite{1992A&A...262...17J}, implying a velocity shift of order 200 km s$^{-1}$ between the narrow and broad components of H$\beta$. One proposed explanation for this is a binary system.

The physical interpretation of the periodicity is uncertain, although its sinusoidal nature suggests that it is kinematic in origin: we consider three possibilities here: (1) The optical flux could be the superposition of thermal emission from the accretion disk and a non-thermal contribution from a precessing jet, and such a model can fit the observed data (see Methods). The expected precession period with a single SMBH is $p \sim 10^{2.8} -10^{6.5}$ years\cite{2005ApJ...635L..17L}, much longer than the observed period. Thus, a binary SMBH origin for the jet precession is more plausible. In this latter case, the jet could precess either as a result of inner disk precession due to the tidal interaction of an inclined secondary SMBH or because of the precession of a circumbinary disk warped by the SMBH binary. (2) Another possibility is a temporary hot spot in the inner region of the accretion disk, but this leads to implausible SMBH mass estimates of $\log(M/M_{\odot}) = 11.4 - 12.2$, depending on the degree of rotation of the SMBH (the largest reported SMBH masses\cite{2011ApJS..194...45S} are of the order of $\log(M/M_{\odot}) \sim 10$). However, with a SMBH binary, periodic mass accretion rates can give rise to an overdense lump in the inner circumbinary accretion disk\cite{2014ApJ...783..134F}. The spectral energy distribution of a circumbinary disk also has a steeper power law\cite{2013ApJ...774..144R} and so accretion variations will have a more noticeable effect at shorter wavelengths. (3) Yet another possibility is a warped disk eclipsing part of the continuum as it precesses, although SMBH binaries are proposed as a possible cause for them\cite{2014MNRAS.441.1408T}. We note as well that light curves for objects known to exhibit these phenomena do not resemble that of PG 1302-102 (see SOM).

If PG 1302-102 were to be described as a binary SMBH pair with a total virial mass of $\log(M/M_{\odot}) \sim 8.5$, the observed period gives an upper limit separation of $\sim$0.01 pc between the pair. This would mean that the system has evolved well into the ``final parsec'' scale. The expectation is that most binary SMBH systems will spend the majority of their lifetime at such separations (0.01 -- 1 pc), in an intermediate phase of evolution between scattering any stars in the nuclear region and gravitational radiation dominance\cite{2008ApJS..174..455B}. 

Further observations could test the different interpretations mentioned above, particularly reverberation mapping to measure the behavior of emission line response to continuum variations, which is expected to be different for alternate explanations\cite{2010ApJ...725..249S}. Continued monitoring by CRTS and other synoptic surveys will obviously track future cycles, and historical photometric data from photographic plate collections may provide more data for previous ones. With decadal baselines, the predicted change in period of the system may be detectable. Future spectroscopic observations can also test whether the line asymmetries vary on binary orbital timescales. Multiwavelength observations should provide more information about the innermost regions of the quasar and the nature of the jet. The relationship between PG 1302-102 and its two nearby companions may also furnish insight into the merger history of this source, particularly as these may contain similarly sized SMBHs to the secondary. Finally, if PG 1302-102 is a SMBH binary, it is a strong candidate for any gravitational wave experiment sensitive to nanohertz frequency waves, such as those using pulsar timing arrays as well as any future space-borne gravitational wave detection mission.

\vspace{35pt}


\bibliography{mybib}{}


\begin{addendum}
 \item This work was supported in part by the NSF grants AST-0909182, IIS-1118031, and AST-1313422. We thank Joseph Scott Stuart, MIT Lincoln Laboratory, for assistance with the LINEAR data. We also thank the staff of the Keck and Palomar Observatories for their help with observations and the CRTS team. Some of the data presented here were obtained at the W.M. Keck Observatory, which is operated as a scientific partnership among the California Institute of Technology, the University of California and NASA. The observatory was made possible by the generous financial support of the W.M. Keck Foundation. The work of D.S. was carried out at Jet Propulsion Laboratory, California Institute of Technology, under a contract with NASA.

\item[Contributions] M.J.G. performed the analysis and wrote the paper. S.G.D. is the PI of the CRTS survey and obtained the Keck spectrum. E.G. obtained and reduced the NIR data and provided the Balmer and Paschen line fits. D.S. reduced the Keck data. A.J.D. is the co-PI of the CRTS survey and provided the CRTS data.
S.L. and E.C. are the PIs of the CSS survey. All authors contributed to the text. 
 
 \item[Competing Interests] The authors declare that they have no
competing financial interests.

 \item[Correspondence] Correspondence and requests for materials
should be addressed to M.J.G.~(email: mjg@caltech.edu).

\end{addendum}

\newpage

\begin{figure}
\centering
\includegraphics[width=0.95\textwidth]{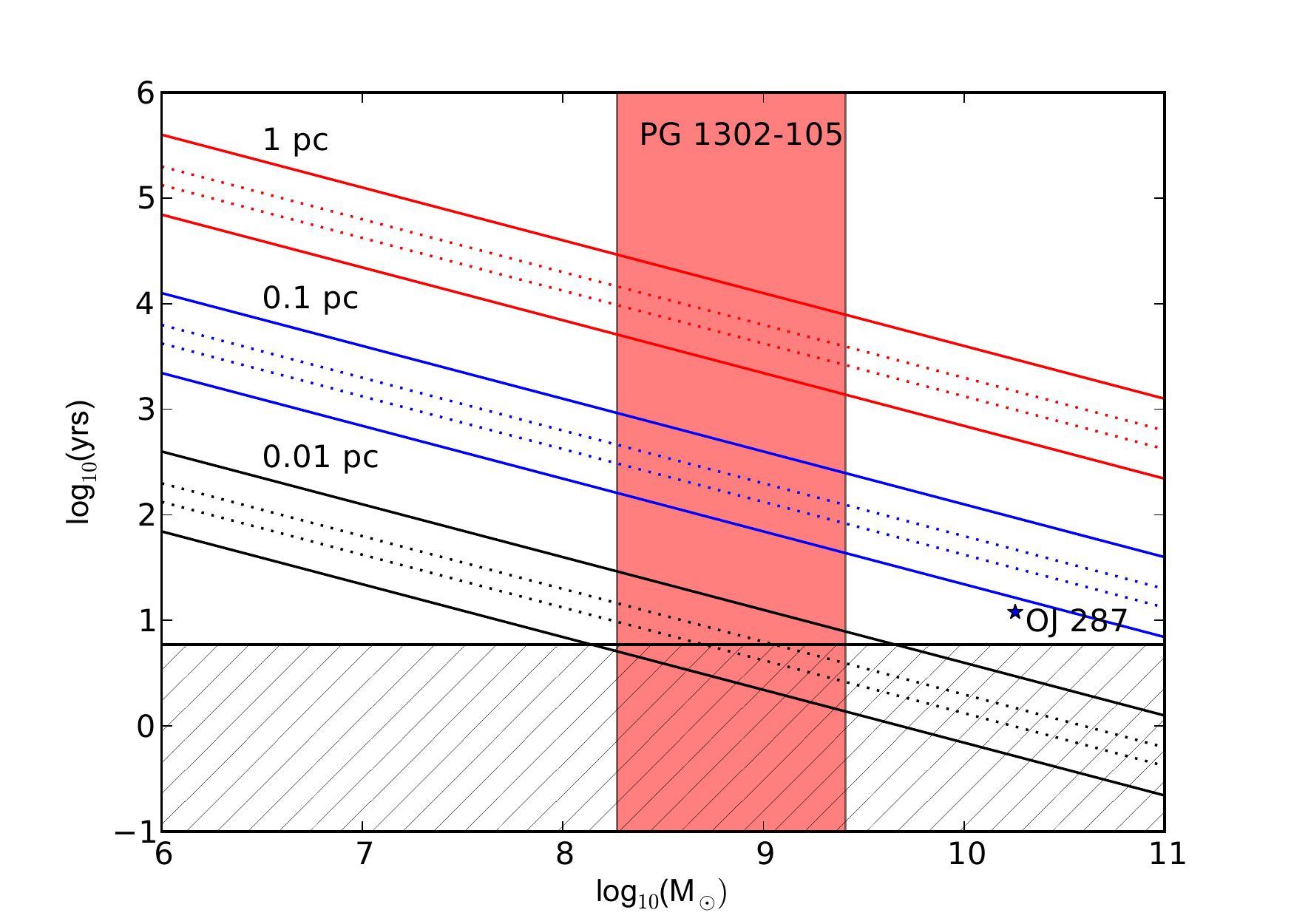}
\caption{{\bf The parameter space of SMBH binary pairs.} The expected orbital periods for SMBH close binary pairs at the specified separations as a function of total black-hole mass. The solid upper line for each separation indicates a $z = 5$ track and the solid lower line  a $z = 0.05$ track whilst the two internal dotted lines show $z = 1.0$ (lower) and $z = 2.0$ (upper) tracks respectively. The hatched region indicates the range over which CRTS has temporal coverage of 1.5 cycles or more of a periodic signal. The pink shaded region shows the region of detection for the best CRTS candidate given the range of virial black-hole masses reported in the literature. Also shown (solid black star) is the location of the best known SMBH binary candidate, OJ 287\cite{2008Natur.452..851V}.}
\end{figure}


\begin{figure}
\centering
\includegraphics[width=0.95\textwidth]{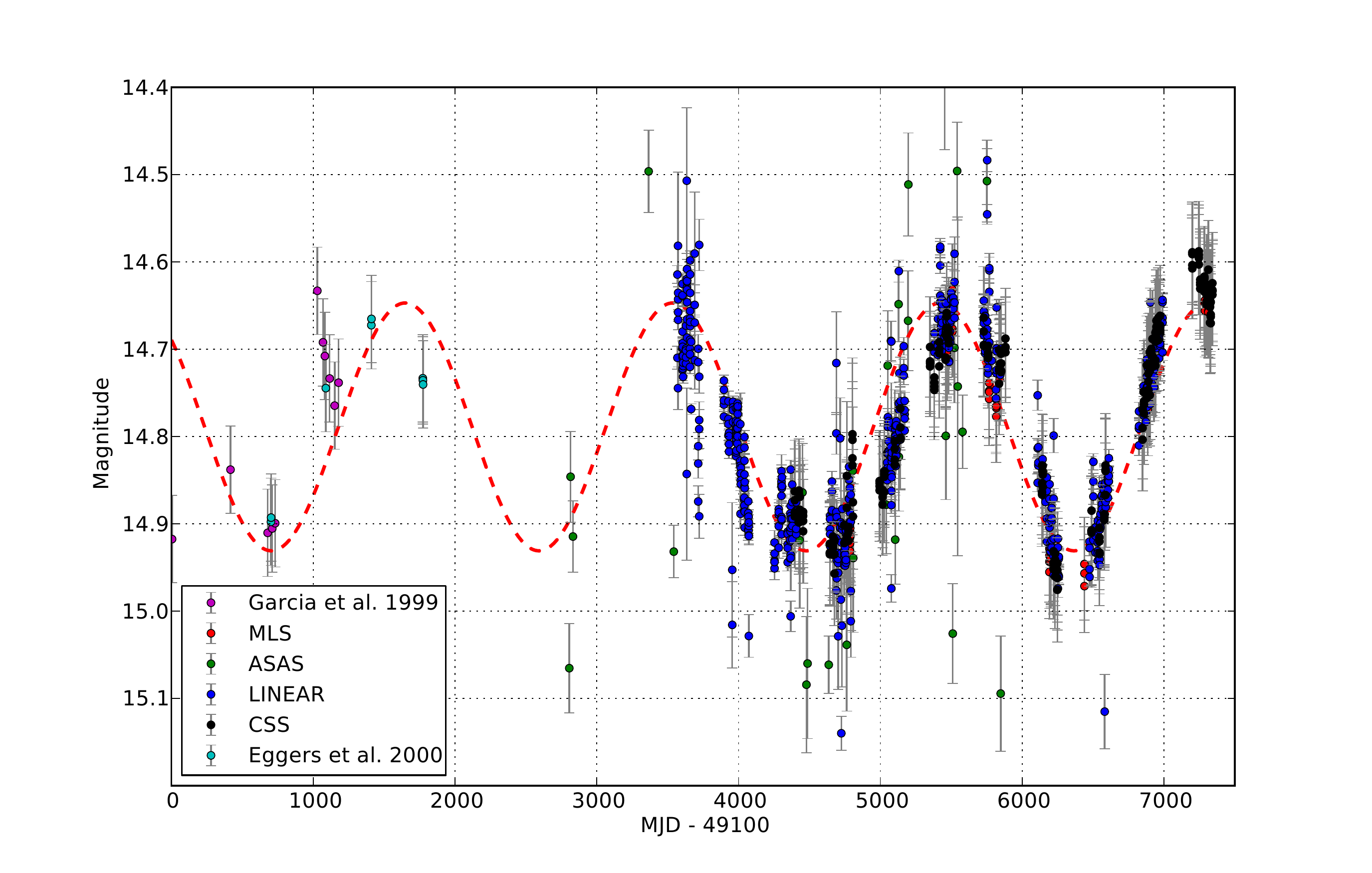}
\caption{{\bf The composite light curve for PG 1302-102 over a period of 7,338 days ($\sim$ 20 years).} The light curve combines data from two CRTS telescopes (CSS and MLS) with historical data from the LINEAR and ASAS surveys, and the literature (see Methods for details). The error bars represent one standard deviation errors on the photometry values. The dashed line indicates a sinusoid with period 1,884 days and amplitude 0.14 mag. The uncertainty in the measured period is 88 days. Note that this does not reflect the expected shape of the periodic waveform which will depend on the physical properties of the system. MJD, modified Julian day.}
\end{figure}


\begin{figure}
\centering
\includegraphics[width=0.95\textwidth]{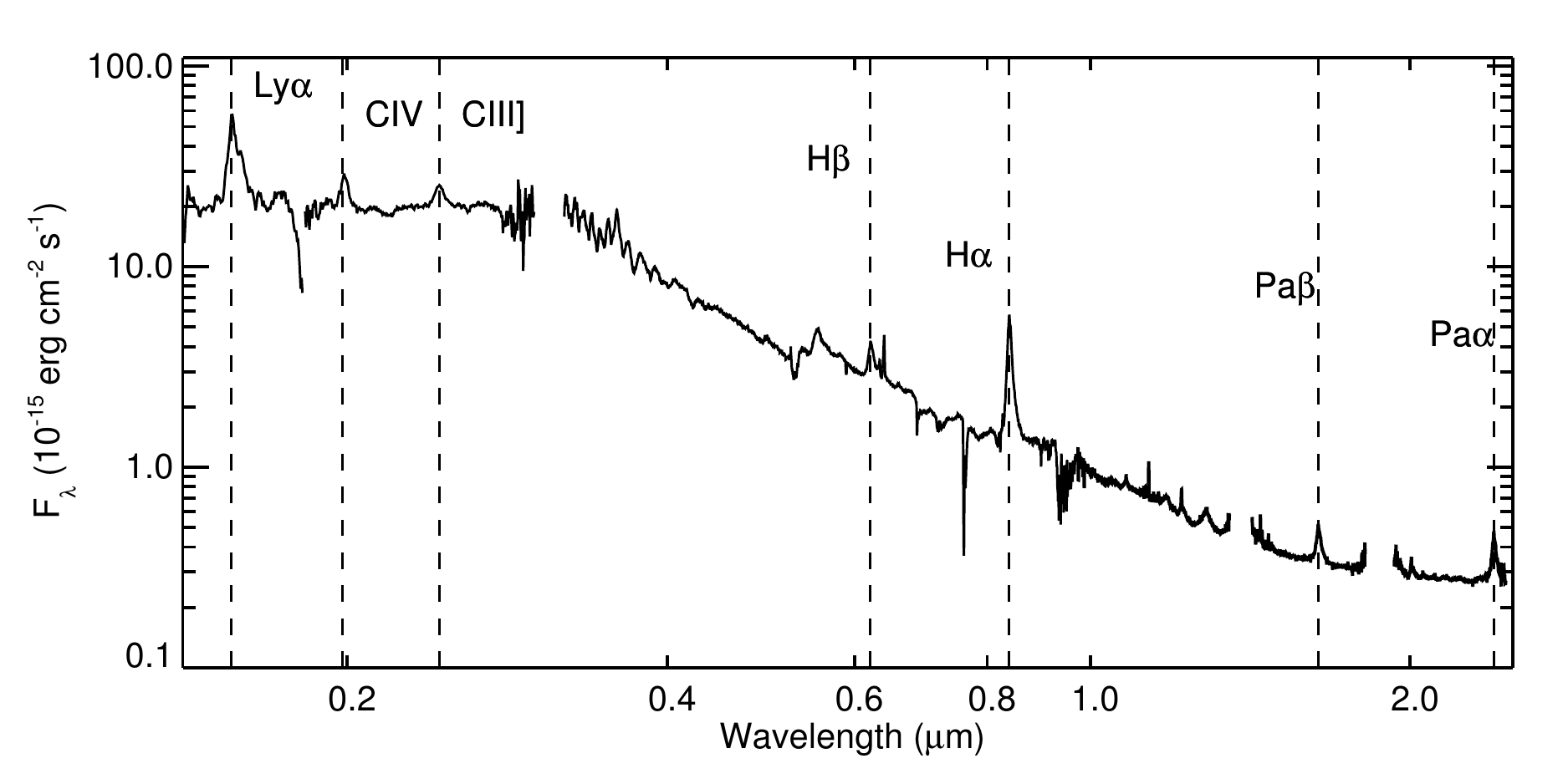}
\caption{{\bf The composite spectrum for PG 1302-102.} This combines an archival GALEX spectrum (ultraviolet) with optical/near-infrared spectra  taken with the Keck and Palomar 200" telescopes in April and June 2014. $F_\lambda$, flux density. The prominent emission lines are indicated. The median flux errors are 5.6 $\times$ 10$^{-16}$ erg s$^{-1}$ cm$^{-2}$ for the GALEX data $(\lambda < 0.3 \mu m)$, 4.5 $\times$ 10$^{-16}$ and 7.6 $\times$ 10$^{-17}$ erg s$^{-1}$ cm$^{-2}$, respectively, for the blue $(0.3 \mu m < \lambda < 0.5 \mu m)$ and red $(0.5 \mu m < \lambda < 0.9 \mu m)$  optical spectra from Palomar, and 4.6 $\times$ 10$^{-18}$ erg s$^{-1}$ cm$^{-2}$ for the Keck near-infrared $(\lambda > 0.9 \mu m)$ spectrum.}
\end{figure}

\begin{figure}
\centering
\includegraphics[width=0.95\textwidth]{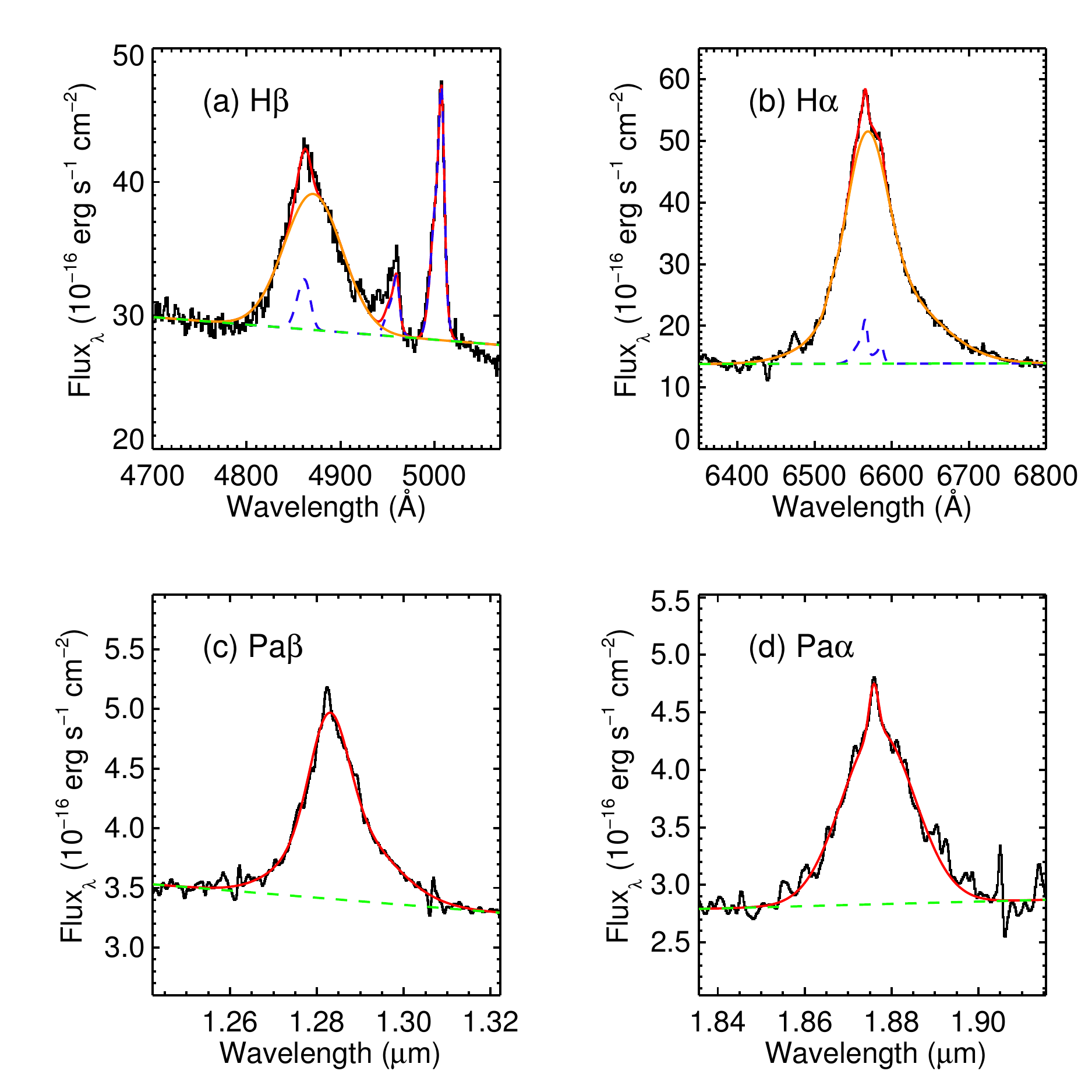}
\caption{{\bf The profiles of the Balmer and Paschen series lines of PG 1302-102.} The data have been modelled with a multi-component line fitting technique (see Methods for details). {\bf a, b} Balmer H$\beta$ (a) and H$\alpha$ (b) have been fit using a narrow component (dashed blue line) and a broad Gaussian (solid orange line). The dashed green line  shows the linear continuum component, and the total fitted profile is shown as a solid red line. H$\beta$ requires a single Gaussian offset from the narrow component but H$\alpha$ requires two components - a central Gaussian plus a red wing. {\bf c,d} The Paschen lines (Pa$\beta$ (c) and Pa$\alpha$ (d)) also show a consistent small asymmetry on the red side.}
\end{figure}

\newpage

\begin{methods}

\section{CRTS data}
The Catalina Real-time Transient Survey (CRTS) makes use of the Catalina Sky Survey which began in 2004, operated by Lunar and Planetary Laboratory at University of Arizona, and uses three telescopes (designated CSS, MLS and SSS) to cover the sky between declination $\delta = -75^{\circ}$ and $+65^{\circ}$  ($\sim$ 80\% of the sky) in order to discover near-Earth objects and potentially hazardous asteroids. The full Catalina surveys data set contains time series for approximately 500 million sources to a limiting magnitude of $V \simeq 20$ with an average of $\sim 250$ observations over a baseline of 9 years per source. CRTS operates an open data policy and the data are publicly available at: http://catalinadata.org.

For photometric calibration we combine observations taken with all three Catalina Sky Survey telescopes since no difference was found between these systems. This is not surprising since each telescope specifically uses the same type of $4{\rm k} \times 4{\rm k}$ CCD camera and observations with all three telescopes are calibrated using the same software pipeline. All Catalina observations are transformed to Johnson $V$ based on 50--100 stars selected as G-type stars using 2MASS\cite{2006AJ....131.1163S} colours.  For bright stars, this photometry provides repeated photometry accurate to $\sim 0.05$\cite{2003DPS....35.3604L}.  However, as the photometry is unfiltered, there are significant variations with object colour.

The 2007 Landolt {\em UBVRI} standard star catalog\cite{2007AJ....133.2502L} provides 109 stars centered near declination $-50^{\circ}$ in the magnitude range $10.4 < V < 15.5$ and in the color index range $-0.33 < (B-V) < 1.66$, while the 2009 Landolt catalog\cite{2009AJ....137.4186L} provides 202 standard stars along the celestial equator in the magnitude range $8.90 < V < 16.30$, and the colour index range $-0.35 < (B - V) < 2.30$, along with 393 standard stars from previous standard star catalogs. In total there are 445 Catalina light curves matching Landolt standards. On average each standard is measured 134 times. A handful of stars that appeared to exhibit significant variability were removed.

Median magnitudes were calculated for each light curve and these were then used to determine the following transformation equations between Johnson $V$ and Catalina $V_{CSS}$:

\begin{eqnarray}\label{tran}
\rm V = V_{CSS} + 0.31 \times (B-V)^2 + 0.04,\\
\rm V = V_{CSS} + 0.91 \times (V-R)^2 + 0.04,\\
\rm V = V_{CSS} + 1.07 \times (V-I)^2 + 0.04.
\end{eqnarray}

\vspace{0.1cm}
The dispersion in the fits to these transformations are 0.059, 0.056 and 0.063 magnitudes, respectively,
for $V < 16$. 

\section{Candidate selection}

We applied the weighted wavelet transform\cite{Foster1996} and the z-transform discrete correlation function (ZDCF)\cite{Alexander2013} to the CRTS light curves of spectroscopically confirmed quasars. Both of these algorithms can detect (quasi-)periodic behaviour in irregularly sampled data. We define the period of the quasar from the largest peak in the ZDCF between the second and third zero-crossings of a Gaussian process model fit to the ZDCF\cite{McQuillan2013}. We have verified that this agrees with periods determined by other methods. The period uncertainty is defined as:

\[ \sigma_P = \frac{1.483 \times MAD}{\sqrt{N-1}} \]

\noindent
where MAD is the median of the absolute deviations from the median of the time intervals between successive peaks in the ZDCF and $N$ is the total number of peaks considered.

We only considered those objects whose wavelet peak significance places them in the top quartile of the data set (in terms of significance). Most kinematically-caused variations should manifest as a (near)-Keplerian signal and so we also use the r.m.s. scatter around an expected sinusoidal waveform (best-fit truncated Fourier series of up to 6 terms) with the ZDCF period to identify those objects most closely exhibiting the expected behavior. We excluded those quasars where the scatter is greater than the $1\sigma$ lower limit on the median absolute deviation of the light curve, i.e., $r.m.s. / mad > 0.67$, since we need to account for the intrinsic variability of the quasar as well.  We also restricted our selection to candidates with temporal coverage of more than 1.5 cycles, assuming the ZDCF period.

\subsection{Simulated data}

For each quasar in our data set, we have generated a simulated light curve assuming that it follows a damped random walk (DRW) model.  Using the actual observation times, $t_i$, we replace the observed magnitudes with those that would be expected under a DRW model. The magnitude $X(t)$ at a given timestep $\Delta t$ from a previous value $X(t - \Delta t)$ is drawn from a Gaussian distribution with mean and variance\cite{2009ApJ...698..895K}:

\[ E(X(t) | X(t - \Delta t)) = e^{-\Delta t / \tau} X(t - \Delta t) + b\tau(1 - e^{-\Delta t / \tau}) \]

\[\mathrm{Var}(X(t) | X(t - \Delta t)) = \frac{\tau \sigma^2}{2} [1 - e^{-2\Delta t / \tau}] \]

\noindent 
We add a Gaussian deviate normalized by the photometric error associated with the magnitude to be replaced at each time $t$ to incorporate measurement uncertainties into the mock light curves. For each light curve, we set $b\tau$ to its median value and use the rest frame DRW fitting functions\cite{MacLeod2010}: 

\[ \log f = A + B \log \left(\frac{\lambda_{RF}}{4000 \AA} \right) + C (M + 23) + D \log \left(\frac{M_{BH}}{10^9 M_{\odot}}\right) \]

\noindent
where $(A,B,C,D) = (-0.51, -0.479, 0.113, 0.18)$ for $f = SF_\infty = \tau\sigma^2/2$ and $(A,B,C,D) = (2.4, 0.17, 0.03, 0.21)$ for $f = \tau$. $M$ is the absolute magnitude of the quasar and $\lambda_{RF}$ is the restframe wavelength of the filter. The mass of the black hole is either the measured virial mass\cite{Shen2011} or is drawn from a Gaussian distribution\cite{MacLeod2010}:

\[
p(\log M_{BH} | M) = \frac{1}{\sqrt{2\pi}\sigma} \exp \left[ - \frac{(\log M_{BH} - \mu)^2}{2\sigma^2}\right]
\] 

\noindent
where $\mu = 2.0 - 0.27 M$ and $\sigma = 0.58 + 0.011 M$. 

\subsection{Statistical significance}

To assess the statistical significance of the detection of PG 1302-102, we generated 1,000 simulated light curves as above. From these we determined the mean weighted wavelet power spectrum as a function of time and frequency and its variance. The dominant signal in the observed WWZ spectrum of PG 1302-102 is then seen to be $\sim40$ times above the corresponding mean DRW value in terms of the expected standard deviation.

We have also performed a periodicity analysis of the light curve of PG 1302-102 using the generalized Lomb-Scargle method which shows a statistically significant peak at the same period identified by the wavelet and autocorrelation analyses. The false alarm probability is $<< 10^{-13}$ -- the $10^{-13}$ level is $P(\omega) = 0.335$ and the observed peak is at $P(\omega) = 0.818$. 

\subsection{Theoretical predictions}

Simple disk models for circumbinary gas and the binary-disk interaction have been used\cite{Haiman2009} to consider the number of SMBH binaries expected in a variety of surveys, assuming that such objects are in the final gravitational wave-dominated phase of coalescence (this equates to separations less than $\sim$0.01 pc for a $10^8 M_\odot$ SMBH binary). This approach has been combined\cite{Volonteri2009} with merger tree assembly models to similarly predict the number of expected SMBH binaries at wider separations where spectral line shifts may be seen (this equates to separations greater than $\sim0.2$ pc for a $10^8 M_\odot$ SMBH binary). The latter shows that in a sample of 10,000 quasars at $z < 0.7$, there should be $\sim$10 objects and this number increases by a factor of $\sim5 - 10$ for $z < 1$. We note, however, that these theoretical arguments are still subject to considerable uncertainties; for example, if the final parsec problem cannot be resolved then there will not be any binaries in the $\sim$0.01 pc regime.

Assuming a limiting magnitude of $V \simeq 20$, a detectable range of orbital periods from 20--300 weeks (spanning both GW- and gas-dominated regimes), a survey sky coverage of 2$\pi$ steradians, and a redshift range of 0.5 -- 4.5, we would expect ~450 SMBH binaries following these approaches. Our finding of 20 candidates from a sample of 240,000 quasars is therefore conservative. 89,000 quasars in our sample also have virial black hole mass estimates\cite{Shen2011} (23\% at $z > 2$) and if we assumed that each of these was a SMBH binary with a separation of 0.01 pc then the CRTS temporal baseline is sufficient to detect 1.5 cycles or more in 63\% of them (including 55\% of the $z > 2$ population). Our search is therefore sensitive to a large fraction of the close SMBH binary population. 

We note that our approach assumes that periodicity associated with SMBH binaries manifests in a Keplerian form. If there is a larger set of non-Keplerian periodic SMBH binaries, either flaring, such as OJ 287, or not, then the 20 objects we have identified may be a small sample of the total close binary SMBH population.

\section{Archival data}

\subsection{LINEAR data\cite{2011AJ....142..190S}}

These were calibrated with pre-release photometry from Pan-STARRS using the $g$, $r$, and $i$ bandpasses.  Comparison stars with instrumental magnitudes (ccd\_mag) between 14 and 17 were selected within 0.1 degrees of PG 1302-102 in LINEAR images. $g-i$ colours were used to compute an $r$-band correction so that a calibration star with $g-i = 0$ has an instrumental magnitude, ccd\_mag = r. Zeropoints for each frame were then derived based on these stars. The reported bandpass for the calibrated magnitudes is therefore approximately $r$. Magnitude errors are computed by SEXTRACTOR, with typical r.m.s. errors between frames of $\sim 0.1$.

\subsection{ASAS data}
The nominal limiting magnitude for ASAS\cite{asas} is $I\sim13$ and so PG 1302-102 is very close to the detection threshold. The low signal-to-noise for such an object is the primary cause of the large degree of scatter seen in ASAS data for this source.

\subsection{Historic data}

Such data for PG 1302 from previous quasar monitoring campaigns is available in the literature\cite{1999MNRAS.309..803G,2000AJ....119..460E}. To put all data on the same photometric scale, offsets were applied to account for differences in the photometric systems used. Region of temporal overlap between a pair of data sets were used to derive offsets so that both data had the same median value. Where no temporal overlap exists, the phased light curve was used to determine the median offset.

Earlier individual photometric observations also exist of PG 1302-102 but the observational errors on these are typically $\sim$ 0.1 mag and so it is difficult to determine whether they agree with the extrapolated behaviour. They also tend to be in different passbands which requires color terms to convert to the $V$-passband to which CRTS is calibrated. However, colour terms for quasars are known to vary (``bluer when brighter") so a constant value cannot be assumed (the quoted $(B-V)$ values for PG 1302-02 have a range of at least 0.2 mag), which introduces an additional error to the transformed magnitude. Such historical data are thus of limited utility.  

\section{Spectroscopic data}

\subsection{Optical}
An optical spectrum was obtained using the Double Spectrograph on the Hale 200-inch Telescope at Palomar Observatory on UT 2014 April 22.  We obtained two 250~s exposures in cloudy conditions using the 1.0'' wide slit, the 5500~\AA\, dichroic, the 600 $\ell\, {\rm mm}^{-1}$ grating on the blue arm ($\lambda_{\rm
blaze} = 4000$~\AA), and the 316 $\ell\, {\rm mm}^{-1}$ grating on the red arm ($\lambda_{\rm blaze} = 7500$~\AA).

On UT 2014 May 26, we obtained additional spectroscopy of PG~1302-102 using the Low Resolution Imaging Spectrometer (LRIS\cite{Oke1995}) on the Keck~I telescope.  We obtained two
300~s exposures in non-photometric conditions, using the 1.5'' slit, the 5600~\AA\, dichroic, the 600 $\ell\, {\rm mm}^{-1}$ grism on the blue arm ($\lambda_{\rm blaze} = 4000$~\AA) and the
400 $\ell\, {\rm mm}^{-1}$ grating on the red arm ($\lambda_{\rm blaze} = 8500$~\AA).

These set-ups provided moderate resolution spectra across the entire optical window, $\lambda \lambda 3100$~\AA - $1\, \mu$m.  The data from both telescopes were reduced using standard procedures and
calibrated using observations of standard stars obtained on the same (non-photometric) nights. 

\subsection{Near-infrared}
We obtained a NIR spectrum of PKS1302-102 with the TripleSpec instrument on the Palomar 200-inch Hale telescope on UT 2014 April 15. Conditions were clear and the seeing was $\sim$1 arcsec.  The source was observed at an airmass of 1.3890 and was observed for four 300-second exposures in an ABBA dither pattern for a total of 20 minutes of on-source exposure.  A spectrum of telluric standard A0V star HD 112304 was obtained immediately following the source spectrum at an airmass difference of 0.15.  We reduced the data using a modified version of the Spextool data reduction package\cite{Cushing2004} which also performs a telluric correction using the standard star spectrum\cite{Vacca2003}. 

Fig.~3 shows the combined optical and near-infrared spectrum, in which we scaled the (non-photometric) optical spectrum by a factor of 2.86 to meet the near-infrared spectrum in their mutually overlapping wavelength region (0.96 - 1.05 $\mu$m).  The change in flux is likely primarily due to a combination of weather variations and slit losses. The Balmer lines, H$\beta$ and H$\alpha$, and Paschen lines, Pa-$\beta$ and Pa-$\alpha$, are marked with vertical dashed lines.  

\section{System parameters}

To estimate the (total) black hole mass for this source, we used the standard method of single epoch virial BH estimation\cite{Shen2013} and adapted relations derived for Paschen lines in the near-infrared\cite{2010ApJ...724..386K}.  To determine the full-width at half maximum (FWHM) of the broad line component of the four lines marked in Figure~4, we applied a multi-component line fitting technique\cite{Greene2004, Glikman2007} in which we model the narrow-line component of the line profiles by first fitting the [OIII] 4959,5007\AA\ lines and fixing the width of a narrow line component in each profile.  For H$\alpha$ we also include the [NII]6548,6583\AA\ doublet, fixed at a flux ratio of 2.96. The broad component can be modelled by up to three Gaussians. If the ratio of chi-squares for successive fits is greater than 0.8, an additional component is added. The NIR spectrum has an error array which is used in the line modelling and parameter estimation (see below). For the optical spectrum, an error array is estimated from the median difference between adjacent pixels (R. White, personal communication). 

The measured values from the 2014 data are: $4,450 \pm 150$ km s$^{-1}$ for H$\beta$, $2,520 \pm 30$ km s$^{-1}$ for Pa-$\alpha$, and $3,200 \pm 20$ km s$^{-1}$ for Pa-$\beta$. Errors on the FWHM were computed using a Monte Carlo approach: the best-fit model for a line was perturbed with a random draw from the error array at each wavelength element and a new fit made. This was repeated 100 times for each line and the standard deviation of the broad line component FWHM used as the error. These values give a (total) mass for the SMBH of $\log(M/M_{\odot}) = 8.8 \pm 0.6$ (H$\beta$), $8.5 \pm 0.1$ (Pa-$\beta$), and $8.4 \pm 0.1$ (Pa-$\alpha$).  Previously reported values of the FWHM of H$\beta$ give a range of estimates for the SMBH mass in the literature using various techniques of $\log(M/M_{\odot}) = 8.3 - 9.4$ so our results are consistent with these. 

We note that the spectral fits of H$\beta$ and Pa-$\beta$ are not perfect. However, the uncertainties in the FWHM and continuum luminosity that this introduces are small compared to the broad range of SMBH mass estimated from the different lines.

\section{Alternate interpretations}

We present further discussion here on the alternate interpretations considered. We note that none of the objects mentioned pass our candidate selection criteria.

\subsection{Jet related}
The optical flux could be the superposition of thermal contribution from the accretion disk with non-thermal contribution produced by the underlying jet. The flux density of the jet (in the optically thin regime) will be boosted in the observer's frame relative to the comoving frame:

\[ S_j(\nu) = S'_j(\nu)\delta(\phi,\gamma)^{p+\alpha} \]

\noindent
where $\alpha$ is the spectral index $(S_j(\nu) \propto \nu^{-\alpha})$ and $p = 2$ for a continuous jet. The jet is precessing with constant angular velocity $\omega$, has an opening angle $\Omega$ and an axis defined by the angles $\phi_0$ (between the jet axis and the line of sight) and $\eta_0$ (the position angle in the plane of the sky):

\[ \sin^2\phi = (\sin\Omega\cos\omega t +\cos\Omega\sin\phi_0\sin\eta_0)^2 + (\sin\Omega\cos\phi_0\sin\omega t + \cos\Omega\sin\phi_0\cos\eta_0)^2
\]

\noindent
Assuming a constant Lorentz factor $\gamma$ for the relativistic bulk motion of the jet, $\gamma = (1-\beta^2)^{-1/2}$, and the Doppler factor is: $\delta = \gamma^{-1}(1-\beta\cos\phi)^{-1}$. Modeling the light curve in this way, we get best-fit parameters of: $\gamma = 5.4 \pm 0.1$, $\Omega = 0.5^\circ \pm 0.1$, $\phi_0 = 5.0^\circ \pm 0.2$, and $\eta_0 = 0.6^\circ \pm 1.4$ (assuming $\alpha = 1.66$).

A number of radio-loud quasars have been reported\cite{Raiteri2001, Fan2002, Kudryavtseva2011} as showing periodic variability in their radio light curves. Whilst a SMBH binary could explain this, a more likely explanation is shock interaction with a helical jet or precession of a jet. However, the optical light curves of  these objects (see Extended Data Fig.~1) do not show the distinctive behavior seen in that of PG 1302-102 suggesting that a different physical mechanism is more likely. We note as well that of the 20 objects in our full sample showing optical periodicity, only 3 are associated with a radio source.

\subsection{Warped accretion disk}

Warped disks have been observed in a handful of AGN\cite{Greenhill2003, Herrnstein2005, Kondratko2008, Kuo2011} and the suggestion here is that as a warp precesses, it could obscure a small amount of continuum emission which would then appear quite regular. Again there is no indication of any periodic behavior in the CRTS light curves available for known objects with warped disks (see Extended Data Fig.~2) similar to that seen in PG 1302-102. 

PG 1302-102 shows a 14\% variation in flux which would suggest that the size of the warp in the disk is quite large. This would also be an orientation-dependent phenomenon and as the source is a blazar, its accretion disk should be oriented close to face-on to us and so any obscuring factor should be limited in effect. We also note that many stellar systems with warped accretion disks are resolvable binary systems.

\end{methods}

\begin{figure}
\centering
\includegraphics[width=0.95\textwidth]{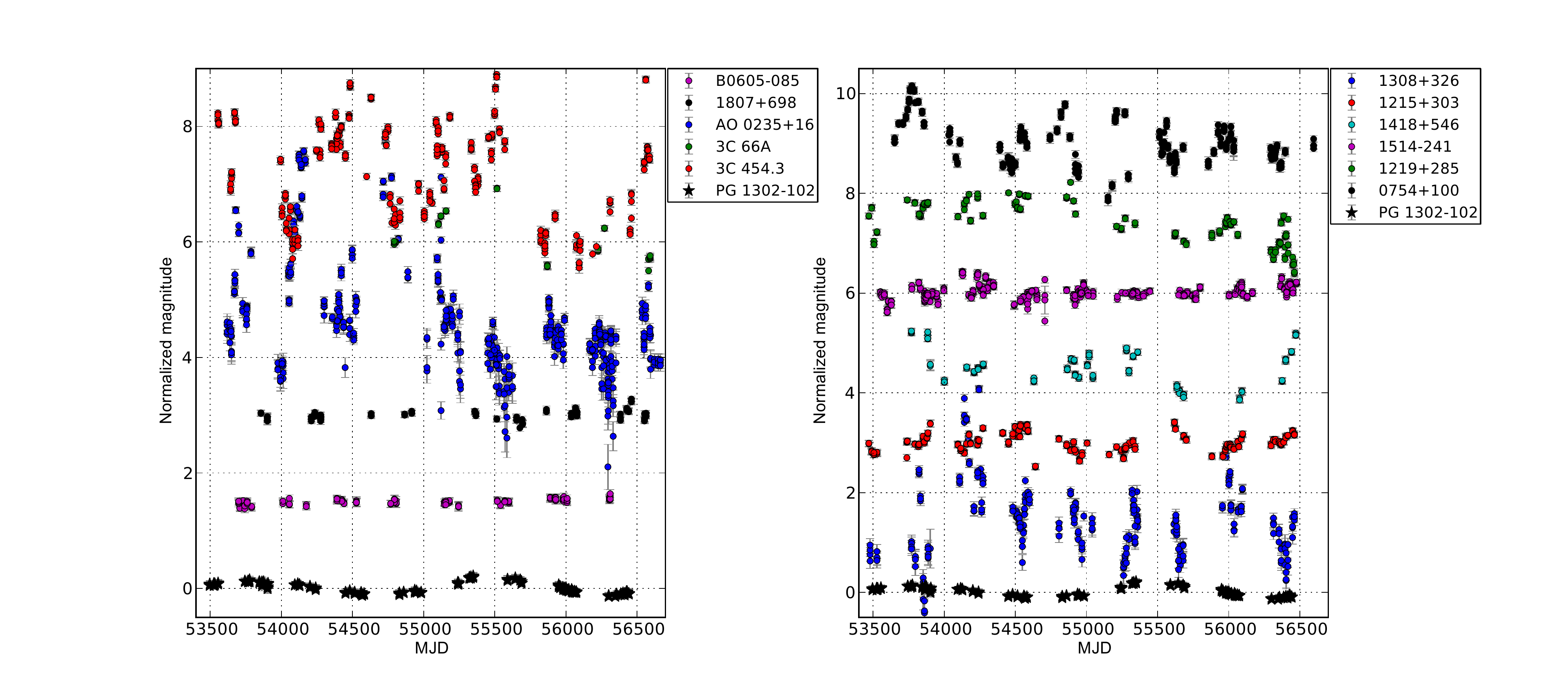}
\caption{{\bf Extended Figure 1  The optical light curves of quasars showing radio periodicity.} Shown are the CRTS light curves for 11 quasars reported\cite{Raiteri2001,Fan2002} to show periodicity in their radio emission. Each light curve has been normalized to zero mean and individual curves are offset by a constant of 1.5 mag from each other. The data are split across two panels for ease of viewing. Error bars shown are standard $1\sigma$ photometric errors. The CRTS light curve of PG 1302-102 (solid black stars) is also shown for comparison.}
\end{figure}

\begin{figure}
\centering
\includegraphics[width=0.95\textwidth]{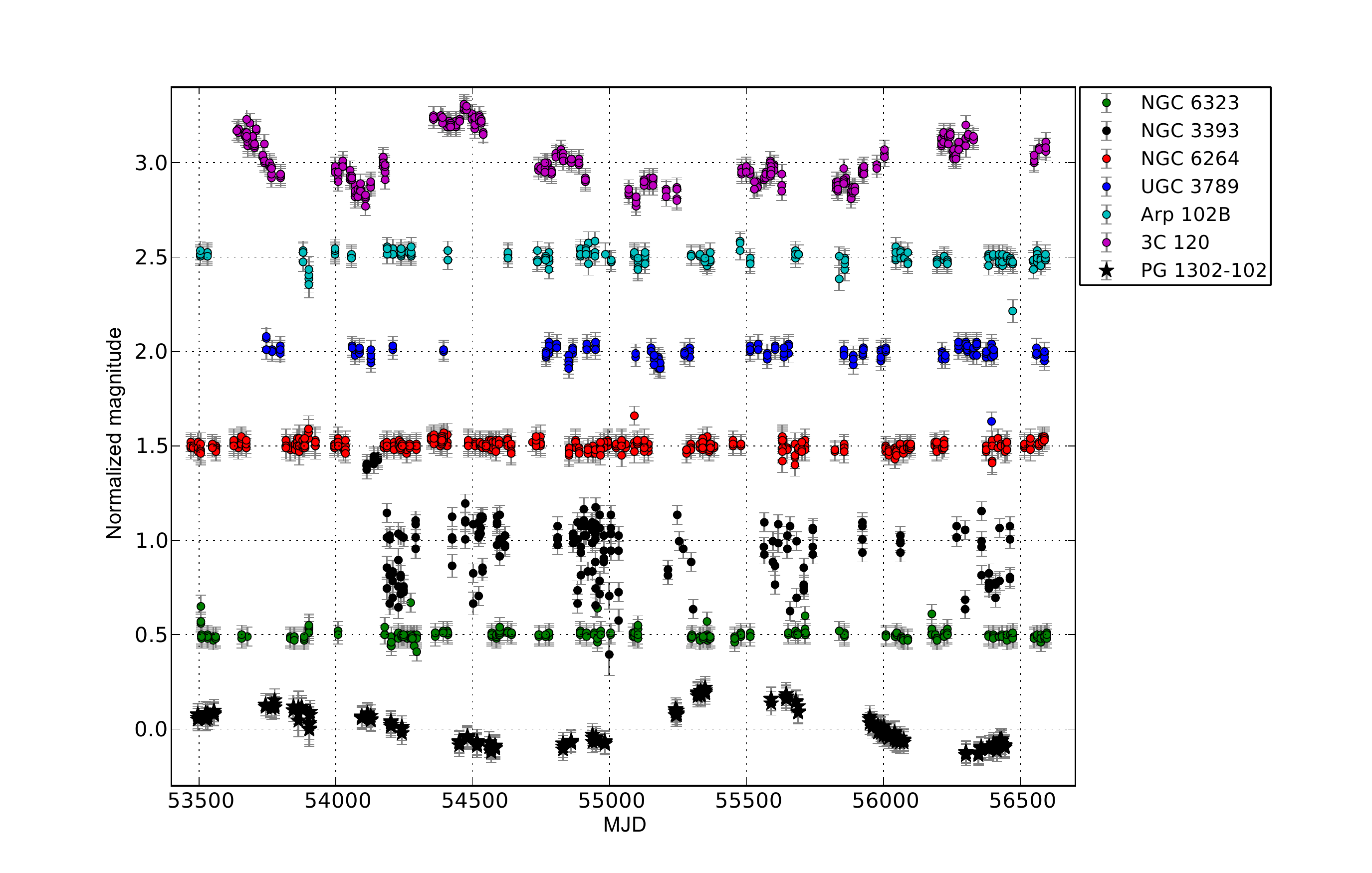}
\caption{{\bf  Extended Figure 2   The optical light curves of quasars with warped accretion disks.} Shown are the CRTS light curves for 6 quasars reported\cite{Greenhill2003, Herrnstein2005, Kondratko2008, Kuo2011} to have warped accretion disks. Each light curve has been normalized to zero mean and individual curves are offset by a constant of 0.5 mag from each other. Error bars shown are standard $1\sigma$ photometric errors. The CRTS light curve of PG 1302-102 (solid black stars) is also shown for comparison.}
\end{figure}

\end{document}